\documentstyle[12pt]{article}

\catcode`@=12
\newcommand{\beq}{\begin{equation}}
\newcommand{\eeq}{\end{equation}}

\newcommand{\bea}{\begin{eqnarray}}
\newcommand{\eea}{\end{eqnarray}}

\def\lsim{\mathrel{\vcenter{\hbox{$<$}\nointerlineskip\hbox{
$\sim$}}}}

\begin{document}
\thispagestyle{empty}
\begin{flushright} June 1999
\end{flushright}
\vspace{0.5in}
\begin{center}
{\Large \bf Models of Neutrino Masses and Baryogenesis\\}
\vspace{1.0in}
{\bf Utpal Sarkar \\}
\vspace{0.2in}
{ \sl Physical Research Laboratory, Ahmedabad 380 009,
India\\}
\vspace{1.0in}
\end{center}
\begin{abstract}
Majorana  masses  of  the  neutrino  implies  lepton  number
violation  and is intimately related to the lepton asymmetry
of  the  universe, which gets related to the baryon asymmetry  of  the
universe in the presence of the sphalerons during the electroweak phase
transition. Assuming that the baryon asymmetry of the universe
is  generated before the electroweak phase transition, it is
possible to dicriminate different classes of models  of
neutrino  masses. While see-saw mechanism  and  the  triplet
higgs mechanism are preferred, the Zee-type radiative models
and  the R-parity breaking models requires additional inputs
to  generate  baryon  asymmetry of the universe  during  the
electroweak phase transition.

\end{abstract}

\newpage
\baselineskip 20pt

\section{Introduction}

Two   important   issues  of lepton   number  violation  are
intimately  related  to  each other.  One  is  the  possible
existence   of  neutrino  Majorana masses, as  evidenced  by
the ongoing excitement generated   by
the      recent     report     of     atmospheric   neutrino
oscillations   \cite{1},  as   well   as    previous   other
indications   of    solar  \cite{2}      and     accelerator
\cite{3}  neutrino oscillations.  The other is one  of   the
very  challenging  question   in cosmology    to    generate
the    baryon  asymmetry of the universe  starting   from  a
symmetric  universe  \cite{kolb}.  Since  the    electroweak
anomalous    processes  breaks  both  the  baryon  and   the
lepton  numbers,   still conserving   the   $(B-L)$  quantum
number,  the baryon asymmetry  of the universe is no  longer
independent  of  the   lepton  number   violation   of   the
universe  \cite{7,fy1,fy2,ht}. If  there  is   very   fast
lepton  number   violation  before  the  electroweak   phase
transition,   then    that    can   erase    the     $(B-L)$
asymmetry  of the universe \cite{fy1} and hence  the  baryon
asymmetry of the universe. On the other  hand, if any lepton
asymmetry  is generated at some high temperature,  that  can
get converted to a baryon asymmetry of  the  universe before
and during  the electroweak phase transition \cite{fy2}.

Lepton number violation is required to give a Majorana  mass
to  the  neutrinos.  Depending on the scale  at  which  this
lepton number  is violated, this interaction may or may  not
satisfy   the   out-of-equilibrium   condition.   If    this
interaction  is  faster  than  the  expansion  rate  of  the
universe, it can erase all lepton asymmetry of the  universe
before the electroweak phase transition. In those models one
then   require  additional  inputs  to  explain  the  baryon
asymmetry  of the universe. On the other hand, in models  of
leptogenesis   the   lepton  number  violating   interaction
required  to  give  Majorana masses  to  the  neutrino  also
satisfy the out-of-equilibrium condition. If there is enough
CP  violation in the leptonic sector \cite{liu},  then  this
can  generate  a  $(B-L)$  asymmetry of  the  universe.  The
anomalous  baryon number violation in the  presence  of  the
sphalerons  will then convert this $(B-L)$  asymmetry  to  a
baryon  asymmetry  of  the universe before  the  electroweak
phase  transition \cite{7,ht}. It is also possible to generate a  baryon
asymmetry  of  the  universe during  the  electroweak  phase
transition  \cite{ewbar},  but  the  condition   that   this
asymmetry  will  not be erased after the  electroweak  phase
transition  gives a strong bound on the mass  of  the  higgs
\cite{higgsbound},  which makes these   less  likely.  As  a
result  leptogenesis  appears  to  be  the  most  attractive
scenario  for generating a baryon asymmetry of the  universe
at  present.  In  this article we shall thus  summarise  the
possibility of leptogenesis in different models of  neutrino
masses.

We  shall first review the original idea of baryogenesis  in
the  context of grand unified theories and show why  $(B-L)$
is  always conserved in the baryon asymmetry thus generated.
Then  we show the relationship between the baryon and lepton
number  in the presence of the sphaleron processes  and  how
$(B+L)$ is washed out. This also implies constraints on  the
lepton number violation and hence on the neutrino masses. At
the  end  we discuss different classes of models of neutrino
masses   which  can  naturally  accomodate
leptogenesis and then summarise.

\section{GUT baryogenesis}

The   subject  of  baryogenesis  originated  when   Sakharov
\cite{sakh} pointed out that for the generation of a  baryon
asymmetry of the universe we need three conditions
\begin{itemize}
\item[(A)] {\it Baryon number violation},
\item[(B)] {\it $C$ and $CP$ violation}, and
\item[(C)] {\it Departure from thermal equilibrium}.
\end{itemize}
It   was  then  realised  that  grand  unified      theories
(GUTs)        satisfies        all       these     criterion
\cite{kolb,gutbar,gutrev}.

The quark-lepton unification implies baryon number violation
in  GUTs.   Since  fermions belong to chiral representation,
$C$   is   maximally   violated.  Departure   from   thermal
equilibrium  was  also naturally satisfied in  these  models
since the scale of unification is sufficiently high, and the
universe  was  expanding very fast in that  epoch.  So,  any
reasonble   GUT   coupling  would   imply   departure   from
equilibrium. Violation of $CP$ was then the
only  crucial point, which  had to be  incorporated in these
theories. However, it was not difficult  to consider some of
the  couplings   to  be complex so  that there  exist   tree
level and one loop diagrams  which could interfere to   give
us  enough  baryon asymmetry in the  decays  of  the   heavy
gauge and higgs  bosons \cite{gutrev}.

This  was  considered to be one  of the major successes   of
GUTs  that  it  can  explain the  baryon  asymmetry  of  the
universe.  After  several years it was  realised  that   the
chiral  nature   of the weak interaction   also  breaks  the
global  baryon   and  lepton numbers in the  standard  model
\cite{hooft}.  Although  both $B$  and  $L$  are  broken,  a
combination $(B-L)$ remains invariant since the  baryon  and
lepton  number  anomalies happens to  be  the  same  in  the
standard model. Since these classical global $(B+L)$  number
symmetry  is broken by quantum effects due  to the  presence
of  the   anomaly, these processes were found   to  be  very
weak   at   the zero temperature. But  at finite temperature
these  $(B+L)$ number violating interactions were found   to
be  very   strong in the presence of some static topological
field  configuration - sphalerons \cite{7}. In fact,  during
the  period  $$10^{12} {\rm GeV} \gg T \gg 10^2 {\rm  GeV}$$
these  interactions  are  so strong  that  in  no  time  the
particles   and  anti-particles  attain  their   equilibrium
distributions.  As a result, since $CPT$  is  conserved  and
hence  the  masses  of the particles and anti-particles  are
same, the number density of baryons becomes same as that  of
the  anti-baryons  and  that will wash  out  any  primordial
$(B+L)$ asymmetry of  the universe. We shall now discuss why
GUT  baryogenesis  always generated only  $(B+L)$  asymmetry
\cite{mamartti},  which  would be erased  by  the  sphaleron
transitions.

In specific GUT scenarios such as $SU(5)$ and $SO(10)$, $(B-
L)$  is  either  a global or a local symmetry  respectively.
Hence the asymmetry generated by the above mechanism is $(B-
L)$  conserving  \cite{gutrev}. When  the  scalar  or  vector
bosons  decay only into fermions, any attempt to generate  a
$(B  -  L)$ asymmetry leads to its large suppression in  all
these  models. We shall now prove that if the decay products
are SM fermions only, this is in fact a generic property  of
any  baryon  asymmetry  generated  by  the  above  described
mechanism. This follows from an operator analysis  analogous
to the one used to show that the minimal scenarios of proton
decay  conserve  $(B-L)$  \cite{wein}.  For  definitness  we
consider  scalars  $X$  and $Y$, but  obviously  the  result
generalizes also to vectors.

Baryogenesis  is possible in GUTs  because there  exist  new
gauge  and Higgs bosons, whose decays violate baryon number.
When  these heavy particles (say $X$) decay into two  quarks
and  into  a quark and an antilepton, the baryon and  lepton
numbers  are  broken  \cite{kolb}. For $CP$  violation  this
mechanism requires two heavy gauge or Higgs bosons, $X$  and
$Y$, each of which should have two decay modes,
\begin{eqnarray}
X \to A + B^* , &~~~{\rm and}~~~& X \to C + D^*\,, \nonumber
\\
Y \to A + C^* , &~~~{\rm and}~~~& Y \to B + D^*\,, \nonumber
\end{eqnarray}
so  that  there exist one-loop vertex corrections  to  these
decays. The  required  $CP$  violation  occurs  due  to   the
interference between tree and loop diagrams. As required  by
the out-of-equilibrium  condition, masses of these particles
must satisfy
\begin{equation}
\Gamma_X  <  H = 1.7 \sqrt{g_*} {T^2 \over M_P} \hskip  .3in
{\rm at}~
T = M_X \,,
\label{hubble}
\end{equation}
where,  $\Gamma_X$ is the decay rate of the  heavy  particle
$X$;  $H$  is  the Hubble constant; $g_*$ is  the  effective
number  of  massless degrees of  freedom; and $M_P$  is  the
Planck scale.

Let  us  start from the Lagrangian giving the decays of  $X$
and $Y$,
\begin{equation}
{\cal L} = f_{x}^{ab} \bar{A} B X + f_{x}^{cd} \bar{C} D X +
f_{y}^{ac} \bar{A} C Y + f_{y}^{bd} \bar{B} D Y\,,
\end{equation}
where  $A,B,C,D$ denote any SM fermion. To obtain a  nonzero
$CP$ violation from the interference between tree and vertex
diagrams,  we require $X$ and $Y$ to be distinct  from  each
other  and to have different decay modes. One can then write
down  all  possible combinations of $A$, $B$, $C$, and  $D$,
with  $X$ and $Y$, and find out the decay modes of  $X$  and
$Y$.   Since  the  out-of-equilibrium  condition   and   the
nonvanishing  of  the absorptive part of the  loop  integral
require  these scalars $X$ and $Y$ to be much  heavier  than
the  fermions, we can integrate them out and write down  the
diagrams in terms of the four-fermion effective operators of
the  SM, as shown in Fig.~1. One can in principle also  have
the  self-energy-type diagrams with the fermions in the loop
for  generating  the $CP$ asymmetry.  In  this  case,  after
integrating out the heavy scalars, the effective diagrams in
terms of the four-fermion operators are exactly the same  as
in  the vertex-correction case, so the conclusions will  not
be changed.

\begin{figure}[htb]
\mbox{}
\vskip 3.75in\relax\noindent\hskip .1in\relax
\includegraphics{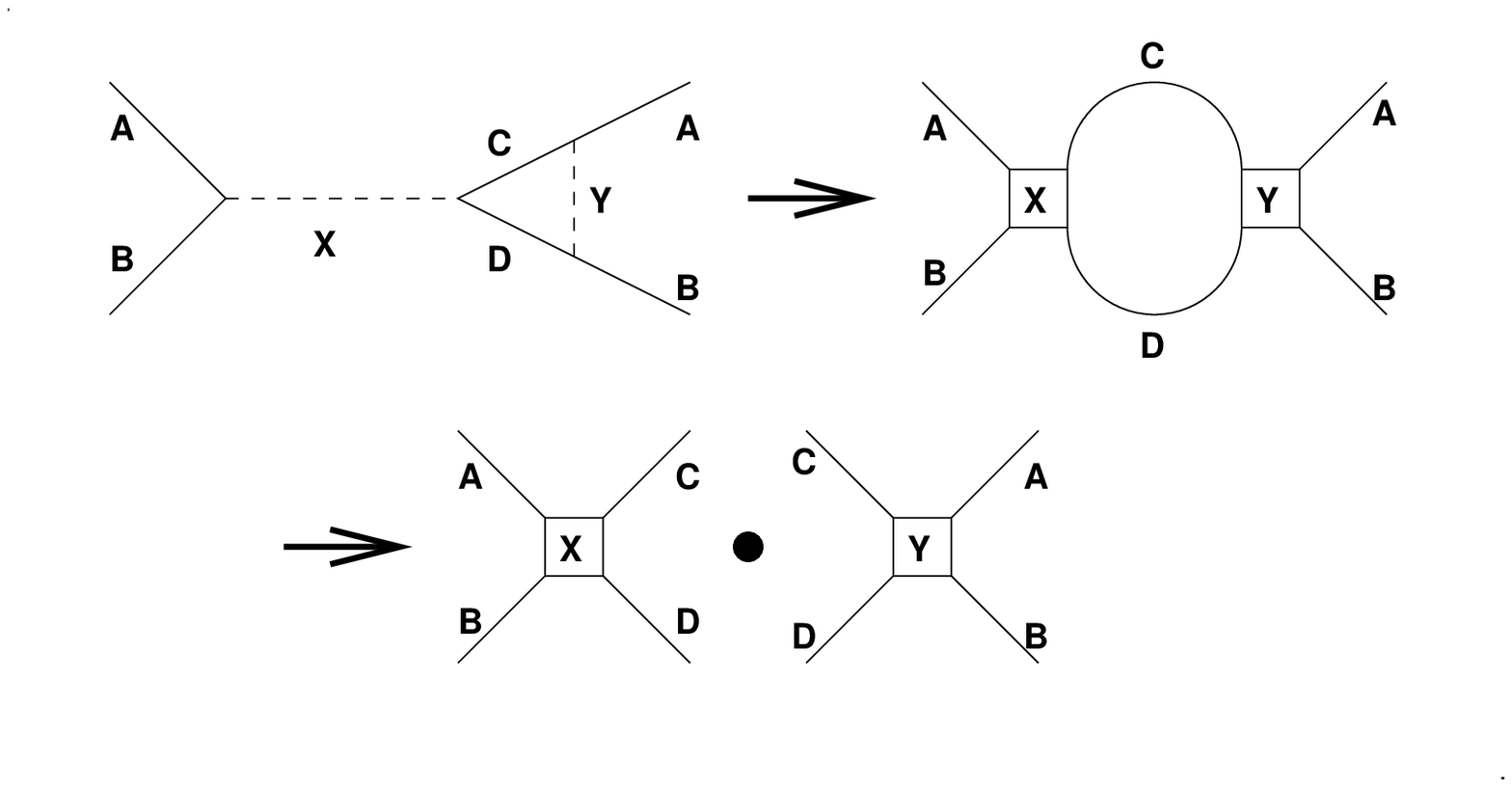}
\vskip -.2in
\caption{ Interference of effective four-fermion operators
which generates
baryon asymmetry.}
\end{figure}

This  simple  but  crucial step allows us  to  use  existing
knowledge  on  SM four-fermion operators for  baryon  number
violation  which  have  been  studied  extensively  in   the
literature  \cite{wein}.   It   was  found  that  all  these
operators conserve $(B-L)$ to the lowest order. Any  $(B-L)$
violating   operator   will  be  suppressed   by   ${\langle
\phi\rangle^2   /  M_{GUT}^2}$  compared  to   the   $(B+L)$
violating operators. In models with an intermediate symmetry
breaking   scale   or  with  new  Higgs  scalars   at   some
intermediate scales, this suppression factor may be softened
a   little,  but  still  strong  enough  to  rule  out   any
possibility  of  generating enough $(B-L)$ asymmetry  of  the
universe. On the other hand, any four-fermion operator which
violates only lepton number requires all the fermions to  be
the  same;  hence  it  cannot  generate  the  required  $CP$
asymmetry. Therefore a $(B-L)$ asymmetry, needed to  survive
the  sphaleron processes, is impossible to generate with the
SM four-fermion operators.

\section{Sphaleron  processes  in  thermal  equilibrium  and
relation between baryon and lepton numbers}

Anomaly  breaks any classical symmetry of the lagrangian  at
the  quantum level. So, all local gauge theories  should  be
free   of   anomalies.  However,  there  may  be   anomalies
corresponding to any global current, which means  that  such
global  symmetries  of the classical lagrangian  are  broken
through  quantum effects. In the standard model  the  baryon
and   lepton   number   global  symmetries   are   anomalous
\cite{hooft}
$$ \delta_\mu j^{\mu 5}_{(B+L)} = 6 [ {\alpha_2 \over 8 \pi}
W_a^{\mu  \nu}  \tilde{W}_{a \mu \nu} +  {\alpha_1  \over  8
\pi}
Y^{\mu \nu} \tilde{Y}_{\mu \nu} ] $$
which  will break the $(B+L)$ symmetry. However, the anomaly
corresponding to the baryon and lepton numbers are same  and
as  a  result  there is no anomaly for the  $(B-L)$  charge.
Because  of  the  anomaly \cite{hooft},  $(B+L)$  is  broken
during  the electroweak phase transition, but their rate  is
very small at zero temperature, since they are suppressed by
quantum   tunnelling  probability,  $\exp[-  {2  \pi   \over
\alpha_2} \nu]$.

At finite temperature, this $(B+L)$ number violation becomes
very   fast   in  the  presence  of  a  non-trivial   static
topological  soliton  configuration, called  the  sphalerons
\cite{7}, and the quantum tunnelling suppression factor is
now replaced by the Boltzmann factor $ \exp[- {V_0 \over  T
}  \nu]$  where the potential or the free energy  $V_0$  is
related to the mass of the sphaleron field. As a result,  at
temperatures between
$
10^{12} GeV > T > 10^2 GeV  \label{per}
$
the  sphaleron  mediated baryon and lepton number  violating
processes  are in equilibrium. For the simplest scenario  of
$\nu = 1$, the sphaleron induced processes are $\Delta  B  =
\Delta L = 3$, given by,
\begin{equation}
|vac> \longrightarrow [u_L u_L d_L e^-_L + c_L c_L s_L \mu^-
_L
+ t_L t_L b_L \tau^-_L] . \label{sph}
\end{equation}
It  can  be  shown  that  any $(B-L)$ asymmetry  before  the
electroweak phase transition will get converted to a  baryon
and lepton asymmetry of the universe, which can be seen from
an analysis of the chemical potential \cite{ht}.

Above  the  electroweak scale, all the  particles  could  be
assumed  to  be  ultrarelativistic. The particle  asymmetry,
{\it   i.e.} the difference  between the number of particles
($n_{+}$) and the number of antiparticles ($n_{-}$)  can  be
given  in  terms of the chemical potential of  the  particle
species $\mu$ (for antiparticles the chemical potential is $-
\mu $) as
\begin{equation}
n_{+}-n_{-}=n_{d}{\frac{gT^{3}}{6}}\left(
{\frac{\mu}{T}}\right),
\end{equation}
where $n_{d}=2$ for bosons  and $n_{d}=1$ for fermions.

In the standard model there are quarks and leptons $q_{iL},
u_{iR},
d_{iR}, l_{iL}$ and $e_{iR}$;
where, $i = 1,2,3$ corresponds to three generations.
In addition, the scalar sector consists of the usual Higgs
doublet
$\phi$,
which breaks the electroweak gauge symmetry $SU(2)_L \times
U(1)_Y$
down to $U(1)_{em}$.
In  Table 1, we presented the relevant
interactions and the corresponding relations between
the chemical potentials.  In the third column we give the
chemical potential
which we eliminate using the given relation.  We start with
chemical
potentials of all the quarks ($\mu _{uL},\mu _{dL},\mu
_{uR},\mu _{dR}$);
leptons ($\mu _{aL},\mu _{\nu aL},\mu _{aR}$,  where
$ a=e,\mu,\tau $); gauge bosons ($\mu _{W}$  for $W^{-}$,
and 0 for all others); and the Higgs scalars ($\mu _{-
}^{\phi  },
\mu _{0}^{\phi}$).

\begin{table}[htb]
\caption {Relations among the chemical potentials}
\begin{center}
\begin{tabular}{||c|c|c||}
\hline \hline
Interactions& $\mu$ relations&$\mu $ eliminated \\
\hline
{$D_{\mu }\phi ^{\dagger}D_{\mu
}\phi $}&{$\mu _{W}=\mu _{-}^{\phi }+\mu _{0}^{\phi
}$}&{$\mu_{-}^{\phi }$}\\
{$\overline{q_{L}}\gamma _{\mu}q_{L}W^{\mu }$}&{$\mu
_{dL}=\mu _{uL}+\mu _{W}$}&{$\mu_{dL}^{{}}$}\\
{$\overline{l_{L}}\gamma _{\mu }l_{L}W^{\mu
}$}&{$\mu_{iL}^{{}}=\mu _{\nu iL}^{{}}+\mu
_{W}$}&{$\mu_{iL}$}\\
{$\overline{q_{L}}u_{R}\phi ^{\dagger
}$}&{$\mu_{uR}=\mu _{0}+\mu_{uL}$}&{$\mu_{uR}^{{}}$}\\
{$\overline{q_{L}}d_{R}\phi $}&{$\mu
_{dR}=-\mu_{0}+\mu _{dL}$}&{$\mu_{dR}$}\\
{$\overline{l_{iL}}e_{iR}\phi
$}&{$\mu _{iR}^{{}}=-\mu_{0}+\mu _{iL}^{{}}$}&{$\mu_{iR}$}\\
\hline \hline
\end{tabular}
\end{center}
\end{table}

The chemical potentials of the neutrinos always enter as a
sum
and for that reason we can consider it as one parameter.
We can then express all the chemical potentials in terms of
the following
independent chemical potentials only,
$ \mu _{0}=\mu _{0}^{\phi };~~\mu _{W};~~\mu _{u}=\mu
_{uL};~~
\mu = \sum_i \mu _{i}= \sum_i \mu _{\nu iL} $.
We can further eliminate one of these four potentials by
making use of the
relation given by the sphaleron processes, $
3\mu _{u}+2\mu _{W}+ \mu =0 $.
We then express the baryon number, lepton numbers and the
electric charge and the hypercharge number densities in
terms of these
independent chemical potentials,
\begin{eqnarray}
&&B =12\mu _{u}+6\mu _{W} ; \hskip 1.6in
L_{i} =3\mu +2\mu _{W}-\mu _{0}  \hskip .4in \nonumber \\
&&Q =24 \mu _{u}+(12+2m)\mu _{0}-(4+2m)\mu _{W} ; \hskip
.2in
Q_{3} =-(10+m)\mu _{W} \hskip .4in \nonumber
\end{eqnarray}
where $m$ is the number of Higgs doublets $\phi$.

At temperatures above the electroweak phase transition,
$T>T_{c}$, both
$<Q>$ and $<Q_{3}>$ must vanish, while below the critical
temperature
$<Q>$ should vanish, but since $SU(2)_L$
is now broken we can consider $\mu_0^\phi =0$ and $Q_3 \neq
0$.
These conditions and the
sphaleron induced $B-L$ conserving, $B+L$ violating
condition will
allow us to write down the baryon asymmetry
in terms of the $B-L$ number density as,
\begin{equation}
B(T>T_c) = \frac{24+4m}{66+13m}~(B-L) \hskip .3in
B(T<T_c) = \frac{32+4m}{98+13m}~(B-L).
\end{equation}
Thus the baryon asymmetry of the universe after
the electroweak phase transition will depend only on the 
primordial $(B-L)$ asymmetry of the universe, while all 
the primordial $(B+L)$ asymmetry will be washed out.

Before proceeding further, we shall briefly discuss what do we mean
when we say that some interaction is fast and that will erase some
asymmetry \cite{kolb,sakh,fry}. In equilibrium the number density of
particles with non-zero charge $Q$
would be same as the antiparticle number density since
the expectation value of the conserved charge vanishes.
A mathematical formulation of this statement
reads that the expectation value of any conserved charge $Q$
is given by,
$$ <Q> = \frac{{\rm Tr} \left[ Q e^{-\beta H}\right]}{{\rm
Tr} \left[
e^{-\beta H}\right]}
$$
and since any conserved charge $Q$ is odd while $H$ is even
under ${ CPT}$ transformation this expectation value vanishes.
So for the generation of the baryon asymmetry of the universe
we have to circumvent this theorem either by including nonzero
chemical potential, or go away from equilibrium or violate
${ CPT}$. In most of the popular models ${ CPT}$
conservation is assumed and one starts with vanishing chemical
potential for all the fields which ensures that the entropy is
maximum in chemical equilibrium. Then to generate the baryon
asymmetry of the universe one needs to satisfy the
out-of-equilibrium condition \cite{kolb,sakh,fry}.

The requirement for the out-of-equilibrium condition may also
be stated in a different way \cite{kolb}. If we assume that the
chemical potential associated with $B$ is zero and ${ CPT}$ is
conserved, then in thermal equilibrium the phase space density
of baryons and antibaryons, given by $[1 + exp(\sqrt{p^2 +
m^2}/kT)]^{-1}$ are identical and hence there cannot be any
baryon asymmetry.

Whether a system is is equilibrium or not can be understood
by solving the Boltzmann equations. But a crude way to put the 
out-of-equilibrium condition is to say that the universe 
expands faster than some interaction rate. For example, 
if some B-violating interaction is slower than the expansion 
rate of the universe, this interaction may
not bring the distribution of baryons and antibaryons of
the universe in equilibrium. In other words, before the
chemical potentials of the two states gets equal, they move
apart from each other. Thus we may state the out-of-equilibrium
condition as
\begin{equation}
\Gamma < \sqrt{1.7 g_*} {T^2 \over M_P}
\end{equation}
where, $\Gamma$ is the interaction rate under discussion,
$g_*$ is the effective number of degrees of freedom available 
at that temperature $T$, and $M_P$ is the Planck scale.

\section{Constraints on neutrino masses}

In the standard model there is no lepton number violation. 
However, one can consider a higher
dimensional effective operator which violates $(B-L)$, given
by
\begin{equation}
L = {2 \over M} l_L l_L \phi \phi + h.c.
\end{equation}
There is no origin of such interactions within the standard
model.
So one expects that some new interaction at some high energy
will give us this effective
interaction at low energy. 

The scale of the new interaction
$M$, which is also the scale of lepton number (and also
$(B-L)$ number) violation, will determine if this interaction
is fast enough to erase all primordial $(B-L)$ asymmetry. 
Since during the same time $(B+L)$ asymmetry is also
washed out by the sphaleron transitions, there will not be
any residual baryon asymmetry of the universe after the
electroweak phase transition. As a result, the survival of the
baryon asymmetry of the universe will then require this
interaction to be slower than the expansion rate of the universe,
\begin{equation}
\Gamma_{L \neq 0} \sim {0.122 \over \pi}{T^3 \over M^2} <
1.7 \sqrt{g_*} {T^2 \over M_P} \hskip .5in {\rm at}~ T \sim
100 GeV
\end{equation}
which gives a bound \cite{fy1} on the lepton number violating 
scale to be, $M > 10^9 GeV$. When the higgs doublets $\phi$ 
acquires a $vev$, the
higher dimensional operator will induce a Majorana mass of
the left-handed neutrinos. This
bound on the heavy scale $M$ will then imply a bound on the
mass of the left-handed neutrinos,
$$ m_\nu < 50 keV  . $$

\begin{figure}[htb]
\mbox{}
\vskip 3.5in\relax\noindent\hskip .6in\relax
\includegraphics{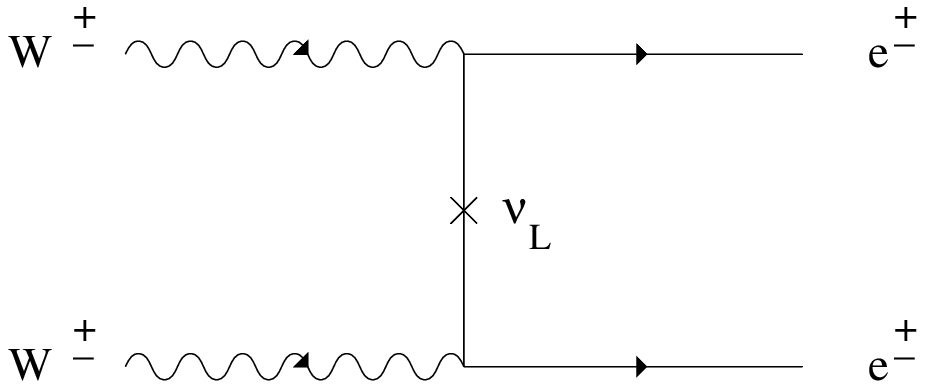}
\vskip -2in
\caption{ Lepton number violating processes $W^{\pm} +
W^{\pm} \to e^{\pm} + e^{\pm}$ mediated by the left handed
Majorana
neutrinos.}
\end{figure}

It is also possible to give a bound on the neutrino mass in
a more general way \cite{us1}. Unless the neutrinos are Dirac
particles \cite{usd}, during the  electroweak  phase  transition 
there will be interactions of the type,
\begin{equation}
W^+ + W^+ \rightarrow e^+_i + e^+_j \hskip .3in {\rm and}
\hskip .3in W^- + W^- \rightarrow e^-_i + e^-_j
\end{equation}
which violate lepton number. These interactions are
mediated by a virtual left-handed neutrino exchange as shown
in figure 2. Here $i$ and $j$ are the generation indices.
Depending on the physical mass
of the left-handed Majorana neutrinos these processes can
wash out any baryon asymmetry between the time when 
the higgs acquires a $vev$ and the $W^\pm$ freeze out, 
{\it i.e.}, between the energy scales 250 GeV and 80 GeV.

The condition that these
processes will be slower than the expansion rate of the
universe,
\begin{equation}
\Gamma (WW \to e_i e_j) = \frac{\alpha_W^2
{(m_\nu)}^2_{ij} T^3}{ m_W^4} < 1.7 \sqrt{g_*}
\frac{T^2}{M_p} \hskip .5in {\rm at} \:\:\: T = M_W
\end{equation}
gives a bound on the Majorana mass of the left-handed
neutrinos
to be,
\begin{equation}
{(m_\nu)}_{ij} < 20 keV .
\end{equation}
This bound is on each and every element
of the mass matrix and not on the physical states.
There are other lepton number violating interactions
like the scattering processes  $\phi + \phi \to l_i +
l_j$ (mediated by a virtual  left-handed  neutrino) 
and decays of  $W^{\pm}$ and the higgs $\phi$,
which also give similar bounds on the left-handed 
neutrino mass.

In some specific models one may give stronger bounds on 
the mass of the neutrinos \cite{lgb1,fisch}. In models with right
handed neutrinos ($N_{Ri},i=e,\mu,\tau$),
the neutrino masses comes from the see-saw
mechanism \cite{seesaw}. The lagrangian for the lepton
sector containing the mass terms of the singlet right handed 
neutrinos $N_i$ and the
Yukawa couplings of these fields with the light leptons is
given by,

\begin{equation}
{\cal L}_{int} = M_i \overline{(N_{Ri})^c} N_{Ri}
h_{\alpha i} \,  \overline{\ell_{L \, \alpha}} \phi \,
N_{Ri}
\end{equation}

\noindent where  $\phi$ is the usual higgs
doublet of the  standard  model;  $l_{L \alpha}$  are the
light leptons,  $h_{\alpha i}$ are the complex  Yukawa
couplings
and $\alpha$ is the generation  index.
Without  loss  of  generality  we work in a basis  in  which
the
Majorana mass matrix of the right handed neutrinos is real
and
diagonal with eigenvalues $M_i$.

Once the higgs doublet $\phi$ acquires a $vev$, the masses
of
the neutrinos in the basis $[\nu_{L \alpha} ~~~ N_{Ri}]$ is
given by,
\begin{equation}
{\cal M}_\nu = \pmatrix{0 & m \cr m & M}
\end{equation}
where, $m \equiv h_{\alpha i} <\phi> $ and $M \equiv M_{ij}$
are $3 \times 3$ matrices. In the limit when all eigenvalues
of $M$ are
much heavier than those of $m$, and the matrix $M$ is not
singular,
this matrix may be block diagonalised. It then gives three
heavy
right handed Majorana neutrinos with masses $\sim M$ and the
Majorana mass matrix of the left-handed neutrinos will be
given
by,
\begin{equation}
m_\nu = m {1 \over M} m^T .
\end{equation}
In this scenario the see-saw masses of the left-handed
neutrinos
explain naturally why
they are so light.

The decay of $N_{Ri}$ into a lepton and an antilepton,
\begin{eqnarray}
  N_{Ri}  &\to& \ell_{jL} + \bar{\phi}, \nonumber \\
   &\to&  {\ell_{jL}}^c + {\phi} .\label{N}
\end{eqnarray}
breaks lepton number. Since the lightest of the right handed
neutrinos
(say $N_1$) will decay at the end, this interaction
($N_1$ decay) should be slow enough so as
not to erase the baryon asymmetry of the universe, which now
implies
\begin{equation}
{|h_{\alpha 1}|^2 \over 16 \pi} M_1 < 1.7 \sqrt{g_*} {T^2
\over M_P}
\hskip .5in {\rm at}~~T = M_1  \label{NN}
\end{equation}
which can then give a very strong bound \cite{lgb1,fisch} on the
mass of the
lightgest of the left-handed neutrinos to be
$$ m_\nu < 4 \times 10^{-3} eV . $$

In models \cite{oth,zee,rpar,triplet,trip}, where the left-handed  
neutrino mass is not related to any heavy  neutrinos  through  
see-saw mechanism, the abovementioned bounds may not be valid.
In addition, there are several  specific
cases even  within the  framework  of see-saw  models (like the 
singular see-saw mechanism where ${\rm det} M = 0$),
where these bounds are not applicable. These bounds are also 
not valid if some global
U(1) symmetry is exactly conserved up to an electroweak
anomaly
\cite{lgb2}. Furthermore, in some very specific models
where a baryon asymmetry of the
universe is generated after the electroweak phase transition
\cite{mm},
or there are some extra baryon number carrying singlets
which decays after the electroweak phase transition \cite{sacha},
it is possible to avoid all the
bounds from constraints of survival of the baryon asymmetry
of the universe.

We shall now discuss similar bounds on the supersymmetric
R-parity violating and Zee type radiative models. Although
the earlier bounds on the see-saw mechanism is not
applicable
when the decays of the right handed neutrinos generate a
lepton asymmetry of the universe, since leptogenesis is not
possible in these R-parity breaking models or the Zee-type
models,
in these models one needs additional inputs to generate a
baryon asymmetry of the universe.

In the R-parity violating models, the unavoidable lepton
number violation at the supersymmetry breaking scale will
erase any primordial $B$ or $L$ or $B - L$ asymmetry
\cite{fisch,7a,mamartti,8}. This is so unless $B-3 L_i$
is conserved \cite{lgb2,10} even after the electroweak phase
transition. This has been pointed out earlier from a general
dimensional analysis, but none of the existing models of
neutrino masses through R-parity violation could accomodate
this symmetry since that cannot allow required neutrino
mixing matrix.
Possible solutions to this problem could be to break R-
parity spontaneously
after the electroweak symmetry breaking \cite{valle}, or to
generate
a baryon asymmetry of the universe in R-parity breaking
scenarios \cite{rpb}, or generate
a baryon asymmetry of the universe  during the electroweak
phase
transition \cite{ewbar}. But these models are incapable of
accomodating
the interesting feature of leptogenesis, namely generating a
baryon asymmetry of the universe from the interaction which
gives a neutrino mass.

Similarly the Zee-type models \cite{zee} considered so
far cannot account for the observed baryon asymmetry of the
universe,
if they have to explain the present neutrino mass spectrum.
Although
the radiative models have the advantage that they can
reproduce
the required maximal mixing naturally, they erase any
primordial
lepton asymmetry of the universe and hence the baryon
asymmetry of
the universe. This severe constraint on the Zee-type models
are
also valid in both supersymmetric and non-supersymmetric
scenarios. It
is not impossible to find an alternative where a baryon
asymmetry
of the universe is generated after it has been washed out by
this
interaction, but that will not be related to the neutrino
mass.

In the MSSM, R-parity of a particle is defined as
\begin{equation}
R \equiv (-1)^{3B + L + 2J},
\end{equation}
where $B$ is its baryon number, $L$ its lepton number, and
$J$ its
spin angular momentum.  Hence the SM particles have
$R = +1$ and their supersymmetric partners have $R = -1$.
Using the
common notation where all chiral superfields are considered
left-handed,
the three families of leptons and quarks are given by
\begin{equation}
L_i = (\nu_i,e_i) \sim (1,2,-1/2), ~~~ e^c_i \sim (1,1,1),
\end{equation}
\begin{equation}
Q_i = (u_i,d_i) \sim (3,2,1/6), ~~~ u^c_i \sim (3^*,1,-2/3),
~~~ d^c_i \sim
(3^*,1,1/3),
\end{equation}
where $i$ is the family index, and the two Higgs doublets
are given by
\begin{equation}
H_1 = (h^0_1,h^-_1) \sim (1,2,-1/2), ~~~ H_2 = (h^+_2,h^0_2)
\sim (1,2,1/2),
\end{equation}
where the $SU(3)_C \times SU(2)_L \times U(1)_Y$ content of
each superfield
is also indicated.  If R-parity is conserved, the
superpotential is
restricted to have only the terms
\begin{equation}
W = \mu H_1 H_2 + f^e_{ij} H_1 L_i e^c_j + f^d_{ij} H_1 Q_i
d^c_j + f^u_{ij}
H_2 Q_i u^c_j.
\end{equation}
If R-parity is violated but not baryon number, then the
superpotential contains the additional terms
\begin{equation}
W' = \epsilon_i L_i H_2 + \lambda_{ijk} L_i L_j e_k^c +
\lambda'_{ijk} L_i Q_j
d_k^c,
\end{equation}
resulting in nonzero neutrino masses either from mixing with
the neutralino mass matrix \cite{6} or in one-loop order \cite{16}.

If lepton-number violating interactions such as
\begin{equation}
L_i + Q_j \to (\tilde d^c_k)^* \to H_1 + Q_l
\end{equation}
are in equilibrium in the early universe, any pre-existing
lepton asymmetry
would be erased.  To make sure that this does not happen,
the following
condition has to be satisfied:
\begin{equation}
{\lambda'^2 T \over 8 \pi} \lsim 1.7 \sqrt{g_*} {T^2
\over M_P} ~~~ {\rm
at}~~T = M_{SUSY}.
\end{equation}
Assuming
that the supersymmetry breaking scale $M_{SUSY}$ is $10^3$
GeV, we find
\begin{equation}
\lambda' \lsim 2 \times 10^{-7},
\end{equation}
which is very much below the typical minimum value of $10^{-
4}$ needed for
radiative neutrino masses \cite{rpar}.  A similar bound was
presented from
dimensional arguments \cite{fisch,7a}.  Larger values of
$\lambda'$ are
allowed if there is a conserved $(B-3L_i)$ symmetry
\cite{lgb2}.  However,
there would be other severe phenomenological restrictions in
that case \cite{10}.  This bound cannot be evaded even if
one uses the bilinear term for neutrino masses instead,
because the induced mixing would introduce trilinear
couplings which violate
lepton number and an effective $\lambda'$ is unavoidable.
This means that although R-parity violation may exist, it
will have very little consequences.  In particular, it will not
contribute significantly to neutrino masses.

In models of radiative neutrino masses \cite{zee}, in
addition to the
suppression due to the $1/16 \pi^2$ factor of each loop,
there is often
another source of suppression due to the Yukawa couplings
involved. In the original Zee model, the SM is
extended to include a charged scalar $\chi^+$ and a second
Higgs doublet.

The relevant terms of the interaction Lagrangian are given
by
\begin{equation}
{\cal L} = \sum_{i < j} f_{ij} (\nu_i e_j - e_i \nu_j)
\chi^+ + \mu
(\phi_1^+ \phi_2^0 - \phi_1^0 \phi_2^+) \chi^- + H.c.,
\end{equation}
where two Higgs doublets are needed or else there would be
no $\phi \phi
\chi$ coupling.  Lepton number is violated in the above by
two units, hence
we expect the realization of an effective dimension-five
operator
$\Lambda^{-1} \phi^0 \phi^0 \nu_i \nu_j$ for naturally small
Majorana
neutrino masses \cite{5}.  This occurs here in one loop and
the elements of
the $3 \times 3$ neutrino mass matrix are given by
\begin{equation}
(m_\nu)_{ij} = f_{ij} (m_i^2 - m_j^2) \left( {\mu v_2 \over
v_1} \right)
F(m_\chi^2,m_{\phi_1}^2),
\end{equation}
where $v_{1,2} \equiv \langle \phi^0_{1,2} \rangle$ and
$m_i$ are the
charged-lepton masses which come from $\phi_1$ but not
$\phi_2$.  The
function $F$ is given by
\begin{equation}
F(m_1^2, m_2^2) = {1 \over 16 \pi^2} {1 \over m_1^2 - m_2^2}
\ln {m_1^2 \over
m_2^2}.
\end{equation}
Since the $m_\tau^2$ terms in Eq.~(11) are likely to be
dominant, this model
has two nearly mass-degenerate neutrinos which mix maximally. 
This is very suitable for explaining the atmospheric
neutrino data \cite{1},
but only in conjunction with the LSND data \cite{3}.  Let
$m_\chi = 1$ TeV,
$m_{\phi_1} = 100$ GeV, $\mu = 100$ GeV, $v_2/v_1 = 1$, and
$f_{\mu \tau} =
f_{e \tau} = 10^{-7}$ to satisfy Eq.~(9), then the
$m_\tau^2$ terms
generate a neutrino mass of 0.0013 eV, which is very much
below the necessary
1 eV or so indicated by the LSND data.  We note that Eq.~(8)
constrains
the combination $f^2/m_\chi$, whereas $m_\nu$ goes like
$f/m_\chi^2$.
Hence neutrino masses would only decrease if we increase
$m_\chi$.  As long
as there is a suppression from $m_\tau^2$ (which comes
of course from
the Yukawa coupling $m_\tau/v_1$), the conflict with
leptogenesis is a real problem.

\section{Models of Leptogenesis}

In the standard model neutrinos are massless.  To make them
massive, there exist four generic mechanisms, namely, the
see-saw mechanism \cite{seesaw}, the triplet higgs mechanism
\cite{triplet,trip}, the radiative mass generation \cite{zee}
and through R-parity violation \cite{6}. All these models
require lepton number violation, which can erase the primordial
baryon asymmetry of the universe. So, the most promising scenario
will be to see if this lepton number violation could be used to
generate a baryon asymmetry of the universe. This is done in
models of leptogenesis \cite{fy2,triplet}. The see-saw
mechanism and the triplet higgs mechanism for neutrino masses are the
two mechanisms, which can accomodate leptogenesis in the
minimal models, which we shall summarise next. Although
one can extend the Zee-type models also or have a complicated
scenario of R-parity breaking models for generating a lepton
asymmetry of the universe \cite{rpb}, 
we shall not discuss them in this article.

\subsection{Leptogenesis with right-handed neutrinos}

To give a small Majorana mass to the left-handed neutrino
through
a see-saw mechanism, right-handed neutrinos were introduced
($N_{Ri},i=e,\mu,\tau$). In these models
neutrino masses come from the see-saw mechanism \cite{seesaw}.
The lagrangian for the lepton sector containing the
mass terms of the singlet right handed neutrinos $N_i$ and
the Yukawa couplings of these fields with the light leptons is
given by eqn (\ref{N}).
Without  loss  of  generality  we work in a basis  in  which
the Majorana mass matrix of the right handed neutrinos is real
and diagonal with eigenvalues $M_i$, and assume $M_3 > M_2 >
M_1$.

Because of the Majorana mass term,
the decay of $N_{Ri}$ into a lepton and an antilepton,
breaks lepton number.
There are two sources of CP violation in this scenario :

\begin{itemize}
\item[$(i)$] vertex type one loop diagrams which interferes with the
tree level diagram given by figure 3. This is similar to the $CP$
violation coming from the penguin diagram in $K-$decays. 

\item[$(ii)$] self energy type one loop diagrams could interfere 
with the tree level diagrams to produce CP violation as shown 
in figure 4. This is similar to the $CP$ violation in $K-\bar{K}$ 
oscillation, entering in the mass matrix of the heavy 
Majorana neutrinos.

\end{itemize}

\newpage
\begin{figure}[htb]
\vskip 2.25in\relax\noindent\hskip -
.3in\relax{\includegraphics{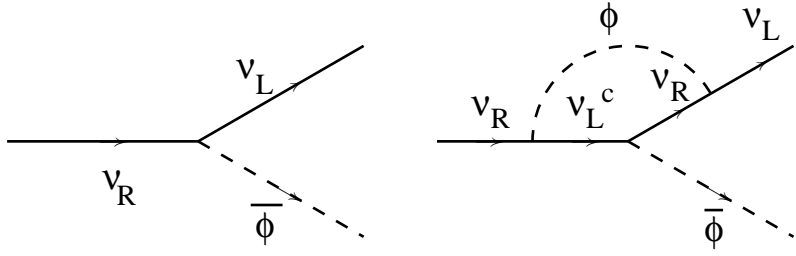}}
\caption{Tree and one loop vertex correction
diagrams contributing to the generation of lepton asymmetry
in models with right handed neutrinos}
\end{figure}

\begin{figure}[htb]
\vskip 2.25in\relax\noindent\hskip -
.3in\relax{\includegraphics{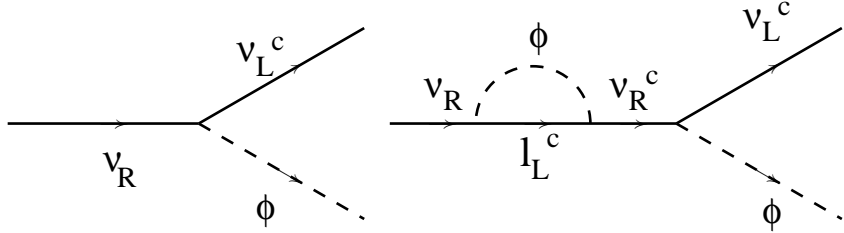}}
\caption{Tree and one loop self energy
diagrams contributing to the generation of lepton asymmetry
in models
with right handed neutrinos}
\end{figure}

In the first paper on leptogenesis \cite{fy1}, the vertex
type diagram was only mentioned. Subsequently, it has been
extensively studied \cite{lepto1} and the amount of $CP$ 
asymmetry is calculated to be,
\begin{equation}
\delta  = - {1 \over 8 \pi} \frac{M_1 M_2}{M_2^2 - M_1^2}
\frac{{\rm Im} [ \sum_\alpha (h_{\alpha 1}^\ast h_{\alpha
2})
 \sum_\beta (h_{\beta 1}^\ast h_{\beta 2}) ] }{ \sum_\alpha
|h_{\alpha 1}|^2}
\end{equation}
In this expression it has been assumed that the main
contribution to the asymmetry comes from the lightest right 
handed neutrino ($N_1$) decay, when the other heavy 
neutrinos have already decayed away.

Initially the self energy diagram was considered for $CP$ violation
as an additional contribution \cite{self}. It was then pointed out
\cite{pas1} that this $CP$ violation enters in the mass matrix as
in the $K-\bar{K}$ oscillation. Before they decay, the right handed
neutrinos were considered to oscillate to an anti-neutrino and since
the rate of $particle \to anti-particle \neq anti-particle \to particle$,
an asymmetry in the right handed neutrino was obtained before they
decay \cite{rv}. As a result, when the two heavy
right handed neutrinos are almost degenerate, {\it i.e.}, the mass
difference is comparable to their width, there may be a resonance
effect which can enhance the $CP$ asymmetry by few orders of
magnitude \cite{pas2}. This effect was then confirmed by other
calculations \cite{pil,rest}, one of which \cite{pil} uses a 
field-theoretic resummation approach \cite{pil1} used earlier to 
treat unstable intermediate states. For large mass difference the
amount of CP asmmetry from the self energy contribution becomes
equal to the vertex correction, which has to be added to get the
final asymmetry. 

Although the $CP$ asymmetry was found to be non-vanishing,
in thermal equilibrium unitarity and $CPT$ would mean that
there is
no asymmetry in the final decay product. However, when the
out-of-equilibrium condition of the heavy neutrinos decay is
considered properly, one could get an asymmetry as expected.
Consider the decays of $K_L$ and
$K_S$. If they were generated in the early universe,
in a short time scale $K_S$ could decay and
recombine, but $K_L$ may not be able to decay or recombine.
As a result in
the decay product there will be an asymmetry in $K$ and
$\bar{K}$
if there is $CP$ violation. In the lepton number violating
two body scattering processes $CP$ violation
in the real intermediate state plays the most crucial role,
which comes since the decay take place away from thermal
equilibrium.

In the case of right handed neutrino decay, the asymmetry is
generated
when the lightest one (say $N_1$) decay. Before its decay,
the pre-existing
lepton asymmetry is washed out by its lepton number
violating interactions.
So the out-of-equilibrium condition now implies that the
lightest
right-handed neutrino should satisfy the out-of-equilibrium
condition when it decays, which is given by,
\begin{equation}
{|h_{\alpha 1}|^2 \over 16 \pi} M_1 < 1.7 \sqrt{g_*} {T^2
\over M_P}
\hskip .5in {\rm at}~~T = M_1
\end{equation}
which gives a bound on the mass of the lightest right-handed
neutrino to be $ m_{N_1} <  10^{7} GeV . $ Finally the
lepton asymmetry and hence a $(B-L)$ asymmetry
generated at this scale gets converted to a baryon asymmetry
of the universe in the presence of sphaleron induced processes.

\subsection{Leptogenesis with triplet higgs}

To give a Majorana mass to the neutrino, one can either 
introduce a right handed neutrino as in the see-saw mechanism,
or else one can introduce two complex $SU(2)_L$ triplet higgs 
\cite{trip,triplet} scalars ($\xi_a \equiv (1,3,-1); a = 1,2$).
The $vev$s of the triplet higgses can give small
Majorana masses to the neutrinos through the interaction
\begin{equation}
f_{ij} [\xi^0 \nu_i \nu_j + \xi^+ (\nu_i l_j + l_i
\nu_j)/\sqrt 2
+ \xi^{++} l_i l_j] + h.c.
\end{equation}
If the triplet higgs acquires a $vev$ and break lepton
number spontaneously,
then there will be Majorons in the problem which is ruled
out by precision
Z--width measurement at LEP. However, in a variant of this
model \cite{triplet}
lepton number is broken explicitly through an interaction
of the triplet with the higgs doublet
\begin{eqnarray}
V &=&  \mu (\bar \xi^0 \phi^0 \phi^0 + \sqrt 2 \xi^- \phi^+
\phi^0 + \xi^{--}
\phi^+ \phi^+) + h.c.
\end{eqnarray}
Let $\langle \phi^0 \rangle = v$ and $\langle \xi^0 \rangle
= u$, then the
conditions for the minimum of the potential relates the
$vev$ of the two scalars by
$ u \simeq {{-\mu v^2} \over M^2}, \label{min} $,
where $M$ is the mass of the triplet higgs scalar and
the neutrino mass matrix becomes $-2 f_{ij} \mu v^2 / M^2 =
2 f_{ij} u$.

In this case the lepton number violation comes from the
decays of the triplet higgs $\xi_a$,
\begin{equation}
\xi_a^{++} \rightarrow \left\{ \begin{array} {l@{\quad}l}
l_i^+ l_j^+ &
(L = -2) \\ \phi^+ \phi^+ & (L = 0) \end{array} \right.
\end{equation}
The coexistence of the above two types of final states
indicates the
nonconservation of lepton number.  On the other hand, any
lepton asymmetry
generated by $\xi_a^{++}$ would be neutralized by the decays
of $\xi_a^{--}$,
unless CP conservation is also violated and the decays are
out of thermal
equilibrium in the early universe. In this case there are no
vertex corrections which can introduce CP violation. The
only source
of CP violation is the self energy diagrams of figure 5.

\begin{figure}[t]
\vskip 2.5in\relax\noindent\hskip -
.5in\relax{\includegraphics{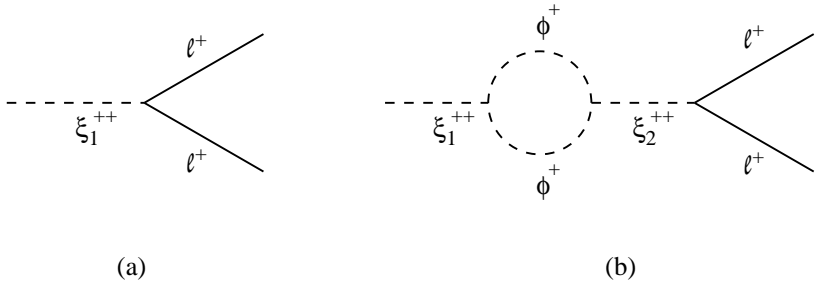}}
\caption{The decay of $\xi_1^{++} \to l^+ l^+$ at tree level
(a) and in
one-loop order (b).  A lepton asymmetry is generated by
their interference
in the triplet higgs model for neutrino masses.}
\end{figure}

If there is only one $\xi$, then the
relative phase between any $f_{ij}$ and $\mu$ can be chosen
real.  Hence
a lepton asymmetry cannot be generated.  With two $\xi$'s,
even if there is
only one lepton family, one relative phase must remain.
As for the possible relative phases
among the $f_{ij}$'s, they cannot generate a lepton
asymmetry because they
all refer to final states of the same lepton number.

In the presence of the one loop diagram, the mass matrix
${M_a}^2$ and
${M_a^*}^2$ becomes different. This implies that the rate of
$\xi_b \to \xi_a$ no longer remains to be same as $\xi_b^*
\to \xi_a^*$.
Since by $CPT$ theorem $\xi_b^* \to \xi_a^* \equiv \xi_a \to
\xi_b$,
what it means is that now $ \Gamma[\xi_a \to \xi_b] \neq
\Gamma[\xi_b \to \xi_a] .$
This is a different kind of CP violation compared to the CP
violation
in models with right handed neutrinos. If we consider that
the $\xi_2$
is heavier than $\xi_1$, then after $\xi_2$ decayed out the
decay of
$\xi_1$ will generate an lepton asymmetry given by,
\begin{equation}
\delta \simeq
{{Im \left[ \mu_1 \mu_2^* \sum_{k,l} f_{1kl} f_{2kl}^*
\right]} \over
{8 \pi^2 (M_1^2 - M_2^2)}} \left[ {{ M_1} \over
\Gamma_1}  \right].
\end{equation}
In this model the out-of-equilibrium condition is satisfied
when
the masses of the triplet higgs scalars are heavier than
$10^{13}$ GeV.

The lepton asymmetry thus generated after the Higgs triplets
decayed away would be the same as the $(B-L)$ asymemtry
before the
electroweak phase transition. During the electroweak phase
transition, the presence of sphaleron fields would relate
this $(B-L)$ asymmetry to the baryon asymmetry of the
universe. The
final baryon asymmetry thus generated can then be given by
the approximate relation $
{n_B \over s} \sim {\delta_2 \over 3 g_* K ({\rm ln}
K)^{0.6}} $. This allows us to obtain a neutrino mass of 
order eV or less, as well as the observed baryon
asymmetry of the universe $n_B/s \sim 10^{-10}$ as desired.

In general, it is not possible to discriminate the see-saw 
mechanism from the triplet higgs mechanism for neutrino
masses. However, in some specific supersymmetric inflationary
models, where the reheat temperature is lower than $!)^{10}$ GeV,
the see-saw mechanism is preferred. On the other hand the 
leptogenesis scenario in the triplet higgs mechanism has 
several nice features, like the absence of the vertex diagrams
or its detectability in the near future in the accelerators
\cite{triplet}. However, in the left-right symmetric models 
both the scenarios are present and can contribute to the
neutrino masses as well as to leptogenesis \cite{lr}.

\section{Summary}

The Majorana masses of the neutrinos implies lepton number violation.
One very important consequence of this lepton number violation in the
early universe is that it can erase any primordial baryon asymmetry of
the universe in the presence of the sphaleron field before the electroweak
phase transition. This gives bound on the mass on the neutrinos. While
a general analysis can give somewhat weak bound, in some specific
models these bounds could be very important. For example, in Zee-type
radiative models or the R-parity breaking supersymmetric models this
is very restrictive. If one attempts to explain the atmospheric neutrino
problem, then these models would wash out all primordial baryon
asymmetry of the universe. This implies that most models of neutrino
masses based on these two scenarios are incomplete and more inputs
are required to explain the present baryon asymmetry of the universe
in these models. On the other hand, in the see-saw mechanism and
the triplet higgs mechanism, the lepton number violation that gives
masses to the neutrinos also generate a lepton asymmetry of
the universe, which then get converted to a baryon asymmetry of the
universe in the presence of the sphaleron field.

\newpage

\bibliographystyle{unsrt}

\end{document}